\documentclass[12pt]{article}
\usepackage{amssymb,amsmath,epsfig}
\usepackage{graphicx}
\allowdisplaybreaks

\begin{document}
\title{\textbf{Impact of Non-metricity and Matter Source on the Geometry
of Anisotropic Spheres}}
\author{M. Zeeshan Gul$^{1,2}$ \thanks{mzeeshangul.math@gmail.com}~,
M. Sharif$^1$ \thanks {msharif.math@pu.edu.pk}~, Adeeba Arooj$^1$
\thanks{aarooj933@gmail.com}~ and
Baiju Dayanandan$^3$ \thanks{baiju@unizwa.edu.om}\\
$^1$ Department of Mathematics and Statistics, The University of Lahore,\\
1-KM Defence Road Lahore-54000, Pakistan.\\
$^2$ Research Center of Astrophysics and Cosmology, Khazar
University,\\ Baku, AZ1096, 41 Mehseti Street, Azerbaijan.\\
$^3$Natural and Medical Sciences Research Center, University of
Nizwa, Oman.}

\date{}
\maketitle

\begin{abstract}
This paper delves into the impact of extended symmetric teleparallel
theory on anisotropic compact stellar structures. The explicit field
equations were formulated by considering a minimum model of this
extended gravity. Basically, the Darmois junction conditions are
used to determine the unknown constants of metric coefficients. We
explore some significant properties of the compact stars under
consideration to check their viable existence in this modified
framework. The Tolman-Oppenheimer-Volkoff equation assess the
equilibrium state of the compact stars. Moreover, the stability
analysis is defined by using methods based on sound speed (related
to how disturbances propagate in the star) and adiabatic index
(related to the thermodynamic behavior of the star). We find that
the proposed compact stars in the $f(\mathbb{Q}, \mathbb{T})$
gravity are physically viable and stable.
\end{abstract}

\section{Introduction}

General theory of relativity (GR) proposed by Albert Einstein define
the concept of gravity as a curvature of spacetime as a result of
mass and energy. Einstein's gravitational theory provides a new
perspective on the fundamental forces in nature. Riemannian geometry
relies on a metric tensor to define distances and angles in a curved
space. While non-Riemannian geometry introduces additional
structures such as non-metricity and torsion to explore more general
geometric properties. Weyl \cite{1} proposed the non-metricity
tensor, which specifies existence of divergence in the metric
tensor. Teleparallel theory includes torsion representing the
gravitational interaction \cite{2}, however symmetric teleparalell
theory includes the non-metricity to define the gravitational
interaction \cite{3}. Xu et al \cite{4} generalized the symmetric
teleparalell theory by include the trace of the energy-momentum
tensor (EMT) in its action, named as $f(\mathbb{Q}, \mathbb{T})$
theory. The theoretical implications and significance of this theory
are currently being explored in cosmological and astrophysical
contexts. As this theory incorporates non-metricity into
gravitational action, it introduces a new gravitational dynamics and
cosmological solutions. There are different forms of modified
theories such as curvature, torsion and non-metricity-based theories
\cite{4}-\cite{25}.

The $f(\mathbb{Q},\mathbb{T})$ theory introduces a specific
connection of geometry and matter, capturing considerable attention
among researchers due to its profound implications in gravitational
physics. The exploration of this theory is motivated by a desire to
understand its theoretical implications and significance in
astrophysical as well as cosmological contexts.  In
$f(\mathbb{Q},\mathbb{T})$ theory, the inclusion of non-metricity
and the presence of matter source enable a more detailed depiction
of gravitational interactions. Non-metricity represents the
deviation from the Levi-Civita connection, which is the connection
compatible with the metric tensor in GR. By incorporating
non-metricity into the gravitational action, this theory introduces
additional degrees of freedom, leading to new gravitational dynamics
and cosmological solutions. Furthermore, the dependence on the trace
of the energy-momentum tensor allows this theory to capture the
effects of matter content on the gravitational field. Thus, the
motivation behind this theory is to consider a more comprehensive
framework for gravitational physics and cosmology, capable of
addressing current observational discrepancies and offering insights
into fundamental questions about the nature of gravity and the
universe.

In recent research, various studies have delved into the
implications of extended symmetric teleparallel theory. Tayde et al
examined the wormhole solutions in this framework \cite{26}. They
also explored the feasibility of traversable wormholes through
strange matter in $f(\mathbb{Q},\mathbb{T})$ background \cite{27}.
Pradhan et al \cite{28} discussed various physical characteristics
of gravastars such as proper length, energy, entropy and surface
energy density in this framework and found a stable gravastar model.
Bourakadi et al \cite{29} conducted a comprehensive study on the
evolution of primordial black holes in this gravity. Loo et al
\cite{30} investigated $f(\mathbb{Q},\mathbb{T})$ theory with small
anisotropy to study the complete cosmic evolution as our universe is
not isotropic since the Planck era. Narawade et al \cite{31} studied
cosmic accelerated expansion in an extended symmetric teleparallel
gravity to understand evolutionary phase of the universe in modified
theory. Tayde et al \cite{32} used conformal motion to analyze
wormhole geometries in this context. Khurana et al \cite{33}
examined that this modified theory presents valuable insights into
the cosmic acceleration. Shukla et al \cite{34} noted that their
findings aligned well with the latest observational datasets,
indicating that the $f(\mathbb{Q},\mathbb{T})$ gravity accurately
describe cosmic dynamics.

Xu et al \cite{35} examined the diverse avenues of research and
potential of $f(\mathbb{Q},\mathbb{T})$ gravity in reshaping the
understanding of cosmic evolution. Arora and Sahoo \cite{36}
examined that the cosmic evolution can be effectively described by
$f(\mathbb{Q},\mathbb{T})$ theory. A deep understanding of early
cosmos in this theoretical framework has been discussed in
\cite{37}. Arora et al \cite{38} analyzed that the extended
symmetric teleparallel gravity address the phenomenon of late-time
cosmic acceleration. The matter bounce scenario in the same
background has been studied in \cite{39}, offering an alternative
cosmological scenario to the Big Bang theory. Godani and Samanta
\cite{40} used several cosmic parameters to explore the cosmic
evolution in the same theoretical framework. Fajardo \cite{41}
highlighted that this extended theory serves as alternative to
standard cosmic model. Arora et al \cite{42} used energy conditions
to study dark universe in the $f(\mathbb{Q},\mathbb{T})$ theory. The
comprehensive study of implications of astrophysical and
cosmological consequences in this modified framework has been
discussed in \cite{43}-\cite{50}

Viable attributes of CSOs yield significant consequences in
alternative theories. Scientists are intrigued by the process of
cosmic object formation and evolution over time. They investigated
diverse enigmatic facets of galaxies, planets, and stars and
asserted that these make up a substantial proportion of the universe
with its own mysterious characteristics. Stars are massive luminous
spheres composed of dust and gases. At star's core, heat energy
released from nuclear fusion opposes the gravitational force
attempting to collapse the star. Stars exhaust their inner fuel that
cause their mass and pressure to decrease and led to gravitational
contraction which cause them to collapse by their own gravity to
form the new compact stellar objects (CSOs) having small radius such
as white dwarfs, neutron stars and black holes. The idea that
neutron stars was first proposed in \cite{51} and their viable
characteristics under different considerations has been explored in
\cite{52}-\cite{57}. Alternative theories and observational
constraints has been examined in \cite{58}-\cite{67}.

Anisotropic fluids account for the directional dependence of
pressure, which can arise from several factors such as nuclear
interactions, rotation and magnetic fields. At high densities, the
interactions between particles can create different pressures in
different directions due to the influence of strong nuclear forces.
Rapid rotation and strong magnetic fields can induce anisotropic
pressure distributions. For instance, a rotating star may experience
greater pressure along its equatorial plane compared to its poles.
Anisotropic models allow for a more flexible and realistic equation
of state that can better describe the thermodynamic properties of
the matter in these stars. In dense astrophysical environments,
matter may undergo phase transitions (e.g., from hadronic matter to
quark-gluon plasma), leading to anisotropic pressure components.
Anisotropic fluids can lead to different stability conditions
compared to isotropic ones. In compact star models, stability
against gravitational collapse and oscillation modes can be better
understood with anisotropic pressures, which might stabilize certain
configurations that would be unstable in isotropic models.
Anisotropic pressure can play a crucial role in avoiding
singularities in the star, allowing for solutions that are
physically viable. The detection of gravitational waves from neutron
star mergers has revealed complex dynamics that can be modeled more
accurately with anisotropic fluid dynamics, reflecting the actual
physical processes at play. Anisotropic models can provide a richer
structure for the field equations, enabling solutions that capture
the complex nature of CSOs. The influence of anisotropy has been
extensively studied in \cite{68}-\cite{73}.

In the study of CSOs, the assumption of isotropic pressure might be
convenient for simplification, but it often fails to capture the
complex dynamics present in these objects. The physical processes
inherent in stellar evolution, including dissipative phenomena and
dynamic equilibrium, inherently lead to the development of pressure
anisotropy, reinforcing the necessity to consider it in our models.
Incorporating pressure anisotropy into the modeling of CSOs is not
merely an academic exercise, but it reflects the underlying physical
realities of stellar dynamics. The inherent dissipative processes
during stellar evolution guarantee that anisotropic pressures will
manifest, shaping the final equilibrium state. Thus, models that
overlook this essential feature risk failing to accurately describe
the behavior of these fascinating astrophysical objects. Herrera
\cite{74} highlighted an important aspect that the pressure
anisotropy is not merely a consequence of specific physical
conditions in CSOs, but is rather an unavoidable outcome of the
stellar evolution process. The consideration of anisotropic fluids
in our model not only aligns with the physical phenomena expected in
CSOs but also accommodates the natural evolution of pressure
anisotropy during the system's history. By adopting anisotropic
fluid models, researchers can better capture the essential
characteristics of CSOs, leading to more accurate predictions of
their behavior such as mass-radius relationships and stability
criteria.

Nashed et al \cite{75} in $f(\mathbb{R})$ gravity and Kumar et al
\cite{76} in $f(\mathbb{R},\mathbb{T})$ theory studied the stability
of CSOs. The charged relativistic structure of CSOs in
$f(\mathbb{G},\mathbb{T})$ theory has been explored in \cite{77}.
Lin and Zhai \cite{78} analyzed the impact of dark source terms on
the geometry of CSOs in $f(\mathbb{Q},\mathbb{T})$ theory. Shamir
and Rashid \cite{79} analyzed the effects of energy density and
pressure components in the interior of CSOs in $f(\mathbb{R})$
theory. The stable anisotropic solution in
$f(\mathbb{R},\mathbb{T}^{2})$ theory has been investigated in
\cite{80}-\cite{82}. The viability and stability of compact
spherical structures with different considerations has been analyzed
in \cite{83}-\cite{85}. Rej and Bhar \cite{86} used observational
data and addition correction terms to determine the behavior of
compact spheres in $f(\mathbb{R},\mathbb{T})$ theory. Das \cite{87}
investigated the role of $f(\mathbb{R},\mathbb{G})$ gravity in the
interior of COSs. Malik et al \cite{88} used Karmarkar condition to
explore the anisotropic CSOs in Ricci-inverse gravity.

This paper follows the given pattern. We first present the field
equations of extended symmetric teleparallel theory in section
\textbf{2} and then derive the explicit expressions for matter
variable by entailing particular $f(\mathbb{Q},\mathbb{T})$ model.
We also identify unknown parameters of the solutions by applying
matching conditions. In section \textbf{3}, we examine the viable
characteristics of considered CSOs through the graphical behavior of
different physical quantities. Section \textbf{4} analyzes their
equilibrium state and stability states. The last section provides
the outcomes of our analysis.

\section{Compact Stellar Objects in $f(\mathbb{Q},\mathbb{T})$ Theory}

Weyl introduced a basic vector field $(h^{\i})$ which allow vector
to alter both its direction and magnitude while being transported
around a closed path. In Weyl geometry, the vector's length along an
infinitesimal path changes as $\delta l= lh_{\i} \delta x^{\i}$
\cite{89} and  the vector's length in a small closed loop is
expressed as
\begin{equation}\label{1}
\delta l= l(\nabla_{\eta}h_{\i}-\nabla_{\i}h_{\eta})\delta s^{\i
\eta}.
\end{equation}
A local length given by $\bar{l}=\i(x)l$ results in the modification
of the field $h_{\i}$ to $\bar{h}_{\i}=h_{\i}+ (\ln\i),_{\i}$, while
elements of the metric tensor undergo changes due to conformal
transformation $\bar{\mathrm{g}}_{\i\eta }=\i^{2}\mathrm{g}_{\i\eta
}$ and $\bar{\mathrm{g}}^{\i\eta }=\i^{-2}\mathrm{g}^{\i\eta }$,
respectively \cite{90}. A semi-metric connection is defined as
\begin{equation}\label{2}
\bar{\Gamma}^{\tau}_{~\i\eta}=\Gamma^{\tau}_{~\i\eta}
+\mathrm{g}_{\i\eta }h^{\tau}-\delta^{\tau}_{~\i}h_{\eta}-
\delta^{\tau}_{~\eta}h_{\i},
\end{equation}
where Christoffel symbol is defined by $\Gamma^{\tau}_{~\i\eta}$. A
gauge covariant derivative can be constructed by assuming that the
symmetric property of $ \bar{\Gamma}^{\tau}_{~\i\eta}$ holds. The
Weyl curvature tensor can then be represented  as
\begin{equation}\label{3}
\bar{\mathbb{C}}_{\i\eta\mu\chi}=\bar{\mathbb{C}} _{(\i\eta
)\mu\chi}+\bar{\mathbb{C}}_{[\i\eta]\mu\chi}.
\end{equation}

After the initial contraction, the Weyl curvature tensor results in
\begin{equation}\label{4}
\bar{\mathbb{C}}^{\i}_{~\eta}=\bar{\mathbb{C}}
^{\mu\i}_{~\mu\eta}=\mathbb{C}^{\i}_{~\eta}
+2h^{\i}h_{\eta}+3\nabla_{\eta}h^{\i}-\nabla_{\i}h^{\eta}
+g^{\i}_{~\eta}(\nabla_{\tau}h^{\tau}-2h_{\tau}h^{\tau}).
\end{equation}
Finally, the Weyl scalar is obtained as
\begin{equation}\label{5}
\bar{\mathbb{C}}=\bar{\mathbb{C}}^{\tau}_{~\tau}=
\mathbb{C}+6(\nabla_{\i}h^{\i}-h_{\i}h^{\i}).
\end{equation}

Weyl-Cartan spaces with torsion provides broaden framework by which
magnitude of a vector is established by a symmetric metric tensor
and the rule for parallel transport is dictated by an asymmetric
connection as $d\varpi^{\i} =
-\varpi^{\tau}{\Gamma}^{\i}_{~\mu\eta}dx^{\eta}$ \cite{91}. The
connection for the Weyl-Cartan geometry is expressed as
\begin{equation}\label{6}
\tilde{\Gamma}^{\tau}_{~\i\eta }={\Gamma}^{\tau}_{~\i\eta}
+\mathbb{W}^{\tau}_{~\i}+\mathbb{L}^{\tau}_{~\i\eta},
\end{equation}
where deformation tensor ($\mathbb{L}^{\tau}_{~\i\eta}$ ) and
contortion tensor ($\mathbb{W}^{\tau}_{~\i\eta }$) are defined as
\begin{eqnarray}\label{7}
\mathbb{L}^{\tau}_{~\i\eta }&=&\frac{1}{2}\mathrm{g}^{\mu\chi}
(\mathbb{Q}_{\eta\i\chi}
+\mathbb{Q}_{\i\eta\chi}-\mathbb{Q}_{\mu\i\eta}),
\\\label{8}
\mathbb{W}^{\tau}_{~\i\eta }&=&\tilde{\Gamma}^{\tau}_{~[\i\eta]}
+\mathrm{g}^{\mu\chi}\mathrm{g}_{\i\nu}
\tilde{\Gamma}^{\nu}_{~[\eta\chi]}+\mathrm{g}^{\mu\chi}
\mathrm{g}_{\eta\nu} \tilde{\Gamma}^{\nu}_{~[\i\chi]}.
\end{eqnarray}
Here
\begin{equation}\label{9}
\mathbb{Q}_{\mu\i\eta}=\nabla_{\tau} \mathrm{g}_{~\i\eta }
=-\partial\mathrm{g}_{\i\eta ,\tau}+\mathrm{g}_{~\eta\chi}
\tilde{\Gamma}^{\chi}_{~\i\mu}
+\mathrm{g}_{~\chi\i}\tilde{\Gamma}^{\chi}_{~\eta\mu},
\end{equation}
and $\tilde{\Gamma}^{\tau}_{~\i\eta }$ is Weyl-Cartan connection. By
using Eq.\eqref{6}, we have
\begin{eqnarray}\label{10}
\tilde{\Gamma}^{\tau}_{~\i\eta }&=&{\Gamma}^{\tau}_{~\i\eta }
+\mathrm{g}_{\i\eta }h^{\tau}
-\delta^{\tau}_{~\i}h_{\eta}-\delta^{\tau}_{~\eta}h_{\i}
+\mathbb{W}^{\tau}_{~\i\eta},
\\\label{11}
\mathbb{W}^{\tau}_{~\i\eta }&=&\mathcal{T}^{\tau}
_{~\i\eta}-\mathrm{g}^{~\mu\chi}\mathrm{g}_{~\nu\i}
\mathcal{T}^{\nu\mu}_{~\chi\eta}-\mathrm{g}^{~\mu\chi}
\mathrm{g}_{~\nu\eta} \mathcal{T}^{\nu\mu}_{~\chi\i},
\end{eqnarray}
with
\begin{equation}\label{12}
\mathcal{T}^{\tau}_{~\i\eta }=\frac{1}{2} (\tilde{\Gamma}^{\tau}
_{~\i\eta }-\tilde{\Gamma}^{\tau}_{~\eta\i}).
\end{equation}
The Weyl-Cartan tensor is expressed as
\begin{equation}\label{13}
\tilde{\mathbb{C}}^{\tau}_{~\i\eta\chi} =\tilde{\Gamma}^{\tau}
_{~\i\chi,\eta} -\tilde{\Gamma}^{\tau}_{~\i\eta,\chi}+\tilde{\Gamma}
^{\tau}_{~\i\chi} \tilde{\Gamma}^{\nu}_{~\mu\eta}-\tilde{\Gamma}
^{\tau}_{~\i\eta} \tilde{\Gamma}^{\nu}_{~\mu\chi}.
\end{equation}
The Weyl-Cartan scalar is obtained by the contraction of this
equation as
\begin{eqnarray}\nonumber
\tilde{\mathbb{C}}&=&\tilde{\mathbb{C}}^{\i\eta} _{~\i\eta}
=\mathbb{C}+6\nabla_{\eta}h^{\eta}-4\nabla_{\eta}
\mathcal{T}^{\eta}-6h_{\eta}h^{\eta}
+8h_{\eta}\mathcal{T}^{\eta}+\mathcal{T}^{\i\mu\eta}
\mathcal{T}_{\i\mu\eta}
\\\label{14}
&+&2\mathcal{T}^{\i\mu\eta}\mathcal{T}_{\eta\mu\i}-4\mathcal{T}^{\eta}\mathcal{T}_{\eta}.
\end{eqnarray}

The elimination of boundary terms in curvature invariant yields
\cite{92}
\begin{equation}\label{15}
\mathcal{I}=\frac{1}{2\kappa} \int \mathrm{g}^{\i\eta
}(\Gamma^{\tau}_{~\chi\i}\Gamma^{\chi}_{~\mu\eta}
-\Gamma^{\tau}_{~\chi\mu}\Gamma^{\chi}_{~\i\eta
})\sqrt{-\mathrm{g}}d^{4}x.
\end{equation}
Based on the assumption of symmetric connection ($
\Gamma^{\tau}_{~\i\eta }=-\mathbb{L}^{\tau}_{~\i\eta })$, we get
\begin{equation}\label{17}
\mathcal{I}=\frac{1}{2\kappa} \int -\mathrm{g}^{\i\eta
}(\mathbb{L}^{\tau}_{~\chi\i}\mathbb{L}^{\chi}_{~\mu\eta}-
\mathbb{L}^{\tau}_{~\chi\mu} \mathbb{L}^{\chi}_{~\i\eta})
\sqrt{-\mathrm{g}}d^{4}x,
\end{equation}
where
\begin{equation}\label{18}
\mathbb{Q}\equiv-\mathrm{g}^{\i\eta}(\mathbb{L}^{\tau}_{~\chi\i}
\mathbb{L}^{\chi}_{~\mu\eta}
-\mathbb{L}^{\tau}_{~\chi\mu}\mathbb{L}^{\chi}_{~\i\eta}),
\end{equation}
with
\begin{equation}\label{19}
\mathbb{L}^{\tau}_{~\chi\i}\equiv-\frac{1}{2} \mathrm{g}^{\mu\nu}
(\nabla_{\i}\mathrm{g}_{\chi\nu}+\nabla_{\chi} \mathrm{g}_{\nu\mu}
-\nabla_{\nu\mu}\mathrm{g}_{\chi\i}).
\end{equation}
By replacing the non-metricity with an arbitrary function in
Eq.(\ref{17}), we obtain
\begin{equation}\label{20}
\mathcal{I}=\int \frac{\sqrt{-\mathrm{g}}}{2\kappa}
f(\mathbb{Q})d^{4}x.
\end{equation}
The integral action with matter part is expressed as
\begin{equation}\label{20b}
\mathcal{I}=\int \frac{\sqrt{-\mathrm{g}}}{2\kappa}
f(\mathbb{Q})d^{4}x+\int \mathcal{L}_{m}\sqrt{-\mathrm{g}}d^{4}x.
\end{equation}
Now, we generalized this action as \cite{4}
\begin{equation}\label{22}
\mathcal{I}=\int \frac{\sqrt{-\mathrm{g}}}{2\kappa}
f(\mathbb{Q},\mathbb{T})d^{4}x+\int
\mathcal{L}_{m}\sqrt{-\mathrm{g}}d^{4}x.
\end{equation}
The superpotential is expressed as
\begin{equation}\label{24}
\mathbb{P}^{\tau}_{~\i\eta }=-\frac{1}{2}\mathbb{L}
^{\tau}_{~\i\eta} +\frac{1}{4}(\mathbb{Q}^{\tau}
-\tilde{\mathbb{Q}}^{\tau})\mathrm{g}_{\i\eta}- \frac{1}{4} \delta
^{\tau}_{(\i\mathbb{Q}_{\eta})},
\end{equation}
where
\begin{eqnarray}\label{23}
\mathbb{Q}_{\tau}\equiv \mathbb{Q}^{~~\i}_{\tau~~\i}, \quad
\tilde{\mathbb{Q}}_{\tau}\equiv \mathbb{Q}^{\i}_{~~\mu\i}.
\end{eqnarray}
The non-metricity is given by
\begin{equation}\label{25}
\mathbb{Q}=-\mathbb{Q}_{\mu\i\eta}\mathbb{P}^{\mu\i\eta}=-\frac{1}{4}
(-\mathbb{Q}^{\mu\eta\chi}
\mathbb{Q}_{\mu\eta\chi}+2\mathbb{Q}^{\mu\eta\chi}
\mathbb{Q}_{\chi\mu\eta}
-2\mathbb{Q}^{\chi}\tilde{\mathbb{Q}}_{\chi}+\mathbb{Q}
^{\chi}\mathbb{Q}_{\chi}).
\end{equation}
The derivation of this equation is given in Appendix \textbf{A}. The
variation of Eq.(\ref{22}) yields
\begin{eqnarray}\nonumber
\delta\mathcal{I}&=&\int \frac{1}{2\kappa}
\delta(f(\mathbb{Q},\mathbb{T})\sqrt{-\mathrm{g}})d^{4}x+\int \delta
(\mathcal{L}_{m}\sqrt{-\mathrm{g}})d^{4}x,
\\\nonumber
&=&\int
\frac{1}{2\kappa}(-\frac{1}{2}f\mathrm{g}_{\i\eta}\sqrt{-\mathrm{g}}\delta
\mathrm{g}^{\i\eta}+f_{\mathbb{Q}}\sqrt{-\mathrm{g}} \delta
\mathbb{Q}+f_{\mathbb{T}}\sqrt{-\mathrm{g}}\delta \mathbb{T}
-\kappa\mathbb{T}_{\i\eta}
\\\label{26}
&\times& \sqrt{-\mathrm{g}}\delta \mathrm{g}^{\i\eta})d^{4}x.
\end{eqnarray}
The calculation of $\delta\mathbb{Q}$ is bestowed in Appendix
\textbf{B}. Moreover, we define
\begin{eqnarray}\label{27}
\mathbb{T}_{\i\eta}\equiv \frac{-2}{\sqrt{-\mathrm{g}}} \frac{\delta
(\sqrt{-\mathrm{g}} \mathcal{L}_{m})}{\delta \mathrm{g}^{\i\eta}},
\quad \Theta_{\i\eta } \equiv \mathrm{g}^{\mu\chi}\frac{\delta
\mathbb{T}_{\mu\chi}}{\delta \mathrm{g}^{\i\eta}},
\end{eqnarray}
which implies that $ \delta \mathbb{T}= \delta (\mathbb{T}_{\i\eta
}\mathrm{g}^{\i\eta}) = (\mathbb{T}_{\i\eta }+
\Theta_{\i\eta})\delta \mathrm{g}^{\i\eta }$. Thus, Eq.\eqref{26}
becomes
\begin{eqnarray}\nonumber
\delta\mathcal{I}&=&\int \frac{1}{2\kappa}\big(\frac{-1}{2}f
\mathrm{g}_{\i\eta}\sqrt{-\mathrm{g}} \delta \mathrm{g}^{\i\eta } +
f_{\mathbb{T}}(\mathbb{T}_{\i\eta }+ \Theta_{\i\eta
})\sqrt{-\mathrm{g}}\delta\mathrm{g}^{\i\eta}
-f_{\mathbb{Q}}\sqrt{-\mathrm{g}}(\mathbb{P}_{\i\mu\chi}
\mathbb{Q}_{\eta}^{~~\mu\chi} \\\label{28} &-&2\mathbb{Q}^{\mu\chi}
_{~~\i} \mathbb{P}_{\mu\chi\eta})\delta \mathrm{g}^{\i\eta
}+2f_{\mathbb{Q}} \sqrt{-\mathrm{g}} \mathbb{P}_{\mu\i\eta}
\nabla^{\tau} \delta \mathrm{g}^{\i\eta}
-\kappa\mathbb{T}_{\i\eta}\sqrt{-\mathrm{g}} \delta
\mathrm{g}^{\i\eta}\big)d^{4}x.
\end{eqnarray}
Through variational principle $(\delta\mathcal{I}=0)$ and
$\kappa=1$, the corresponding field equations are obtain as
\begin{eqnarray}\nonumber
\mathbb{T}_{\i\eta}&=& \frac{-2}{\sqrt{-\mathrm{\mathrm{g}}}}
\nabla_{\tau} (f_{\mathbb{Q}}\sqrt{-\mathrm{g}}
\mathbb{P}^{\tau}_{~\i\eta })- \frac{1}{2} f \mathrm{g}_{\i\eta} +
f_{\mathbb{T}} (\mathbb{T}_{\i\eta} + \Theta_{\i\eta})
-f_{\mathbb{Q}} (\mathbb{P}_{\i\mu\chi}
\mathbb{Q}_{~~\eta}^{\mu\chi}
\\\label{29}
&-&2\mathbb{Q}^{\mu\chi}_{~~\i} \mathbb{P}_{\mu\chi\eta}),
\end{eqnarray}
where $f_{\mathbb{T}}=\frac{\partial f}{\partial \mathbb{T}}$ and
$f_{\mathbb{Q}}=\frac{\partial f}{\partial \mathbb{Q}}$. The these
equations can help to better comprehend the behavior of gravity in
this modified framework.

We consider interior region as
\begin{equation}\label{7a}
ds^{2}=dt^{2}e^{\xi(r)}-dr^{2}e^{\eta(r)}-r^{2}d\Omega^{2},
\end{equation}
where $d\Omega^{2}=d\theta^{2}+\sin^{2}\theta d\phi^{2}$. The EMT
demonstrates internal configuration of matter and energy in the
system. We assume matter configuration as
\begin{equation}\label{8a}
\mathbb{T}_{\tau\upsilon}=\mathbb{U}_{\tau}\mathbb{U}_{\upsilon}
\varrho + \mathbb{V}_{\tau}\mathbb{V}_{\upsilon}
p_{r}-p_{t}g_{\tau\upsilon} +
\mathbb{U}_{\tau}\mathbb{U}_{\upsilon}p_{t} -
\mathbb{V}_{\tau}\mathbb{V}_{\upsilon}p_{t}.
\end{equation}
We consider $\mathbb{L}_{m}=-\frac{p_r+2p_t}{3}$ \cite{93}, because
this expression has been explored in different theoretical contexts
and effectively elucidates the occurrences connected to non-uniform
material configurations in relation to dense celestial objects. The
resulting field equations are
\begin{eqnarray}\nonumber
\varrho&=&\frac{1}{2r^{2}e^{\eta}}\big(2rf_{\mathbb{Q}
\mathbb{Q}}(e^{\eta}-1)\mathbb{Q}'
+f_{\mathbb{Q}}\big((e^{\eta}-1)(2+r\xi')+(e^{\eta}+1)r\eta' \big)
\\\label{9a}
&+&fr^{2}e^{\eta}\big)-\frac{1}{3}f_{\mathbb{T}}(3\varrho+p_{r}+2p_{t}),
\\\nonumber
p_{r}&=&\frac{-1}{2r^{2}e^{\eta}}\big(2rf_{\mathbb{Q}
\mathbb{Q}}(e^{\eta}-1)\mathbb{Q}'
+f_{\mathbb{Q}}\big((e^{\eta}-1)(2+r\xi'+r\eta')-2r\xi'\big)
\\\label{10a}
&+&fr^{2}e^{\eta}\big)+\frac{2}{3}f_{\mathbb{T}}(p_{t}-p_{r}),
\\\nonumber
p_{t}&=&\frac{-1}{4re^{\eta}}\big(-2r\mathbb{Q}'\xi'f_{\mathbb{Q}
\mathbb{Q}} +f_{\mathbb{Q}}\big(2\xi'(e^{\eta}-2)-r\xi'^{2}
+\eta'(2e^{\eta}+r\xi')
\\\label{11a}
&-&2r\xi''\big)+2fre^{\eta}\big)+\frac{1}{3}f_{\mathbb{T}}
(p_{r}-p_{t}).
\end{eqnarray}

The field equations cannot be solved due to presence of multivariate
functions and derivative terms. To reduce complexity in the field
equations, we consider
\begin{eqnarray}\label{12a}
f(\mathbb{Q},\mathbb{T})=h\mathbb{Q}+k\mathbb{T}.
\end{eqnarray}
We consider the values of model parameters as $h=2$ and $k=3$ in all
graphical analysis. The choice of this functional form is motivated
by several physical considerations beyond the simplicity of analysis
\cite{94}. This form allows to study the individual impacts of
geometric modifications and matter fields on the dynamics of the
universe. The additional term involving $\mathbb{T}$ is used to
address various cosmological observations such as the cosmic
expansion without necessarily invoking dark energy. The parameter
$k$ is adjusted to fit observational data, providing a flexible
framework for cosmological modeling. The considered functional form
avoids the introduction of higher-order derivatives in the field
equations, which lead to ghost degrees of freedom (unstable modes)
in the theory. This is a common issue in many extended theories of
gravity and this form helps to maintain the stability of the theory.
While simplicity of analysis is an important consideration, it is
also physically motivated by the need for tractable models that can
be effectively compared with observational data. The chosen form
leads to field equations that are simpler to solve numerically and
analytically, making it possible to explore a wide range of
scenarios and their implications for astrophysics and cosmology.

The explicit field equations are
\begin{eqnarray}\nonumber
\varrho&=&\frac{h e^{-\eta}}{12r^2(2h^{2}+k-1)}\big(
k(2r(-\eta'(r\xi'+2)+2r\xi''+\xi'(r\xi'+4))-4e^{\eta}
\\\label{13a}
&+&4)+3k r(\xi'(4-r \eta'+r \xi')+2r\xi'')+12 (k-1)(r
\eta'+e^{\eta}-1)\big),
\\\nonumber
p_{r}&=&\frac{h e^{-\eta}}{12r^2(2k^{2}+k-1)}\big(
2k\big(r\eta'(r\xi'+2)+2(e^{\eta}-1)-r(2r\xi''+\xi'(r\xi'
\\\nonumber
&+&4))\big)+3\big(r\big(k \eta'(r \xi'+4)-2k r \xi''-\xi'(-4k+k r
\xi'+4)\big)-4(k-1)
\\\label{14a}
&\times&(e^{\eta}-1)\big)\big),
\\\nonumber
p_{t}&=&\frac{h e^{-\eta}}{12r^2(2k^{2}+k-1)}\big(
2k\big(r\eta'(r\xi'+2)+2(e^{\eta}-1)-r(2r\xi''+\xi'
\\\nonumber
&\times&(r\xi'+4))\big)+3\big(r
\big(2(k-1)r\xi''-((k-1)r\xi'-2)(\eta'-\xi')\big)
\\\label{15a}
&+&4k(e^{\eta}-1)\big)\big).
\end{eqnarray}
We consider viable non-singular solution with arbitrary constants
$(a,b,c)$ as \cite{95}
\begin{eqnarray}\label{16a}
e^{\xi(r)}=a\big(\frac{r^2}{b}+1\big),\quad
e^{\eta(r)}=\frac{\frac{2
r^2}{b}+1}{\big(\frac{r^2}{b}+1\big)\big(1-\frac{r^2}{c}\big)}.
\end{eqnarray}
The viable non-singular solutions describe the gravitational field
around CSOs. The constants of the solutions offer crucial insight on
the viable characteristics of CSOs, facilitating the comprehension
of their attributes. These solutions provide a description of energy
distribution and the gravitational redshift effects close to the
surface of CSOs. The presence of integration constants may determine
the curial cosmic characteristics such as event horizons and
singularities. Hence, the integration constants provide a platform
for investigating the gravitational consequences and structural
characteristics of dense CSOs.

Now, we use the matching criteria between the inner and exterior
spacetimes at the boundary to ascertain the values of the
unidentified constants. More precisely, we employ boundary
conditions that are obtained from the junction conditions. These
constraints guarantee the smoothness of the metric potentials and
their derivatives at the boundary. We obtain numerical values of the
unknown constants by  summarizing the resulting system of equations.
The determined constants have significance impact for the behavior
and attributes of CSOs. The precise values of these constants have a
direct influence on the structural and dynamic characteristics of
CSOs.

Let us assume the external geometry of the CSOs as
\begin{eqnarray}\label{17a}
ds^{2}_{+}=dt^{2}\aleph-d r^{2}\aleph^{-1}-r^{2}d\Omega^{2},
\end{eqnarray}
where $\aleph=\left(1-\frac{2m}{r}\right)$. The metric coefficients
exhibit continuity at the surface boundary $(r=R)$ as
\begin{eqnarray}\nonumber
g_{tt}&=&a\big(\frac{R^{2}}{b}+1\big)=\aleph,
\\\nonumber
g_{rr}&=&\frac{\frac{2
R^2}{b}+1}{\big(\frac{R^2}{b}+1\big)\big(1-\frac{R^2}{c}\big)}=\aleph^{-1}
\\\nonumber
g_{tt,r}&=&\frac{2aR}{b}=\frac{m}{R^{2}}.
\end{eqnarray}
Solving these equations, we have
\begin{eqnarray}\label{18a}
a=1-\frac{3m}{R}, \quad b=\frac{R^{3}-3mR^{2}}{m},\quad
c=\frac{R^{3}}{m}.
\end{eqnarray}
These constants are crucial for understanding the internal structure
of the CSOs. Table \textbf{1} shows the numerical values of these
constants corresponding to observed value of mass and radius for
considered CSOs under consideration. Figure \textbf{1} demonstrates
that the metric coefficients show regularity and increasing trend,
indicating non-singular spacetime. In the graphs, we use the colors
gray, pink, green and brown for CSOs given in Table 1, respectively.
\begin{table}\caption{Values of unknown constants.}
\begin{center}
\begin{tabular}{|c|c|c|c|c|c|}
\hline Stellar objects & $m_{\odot}$ & ${R}(km)$ & a & b & c
\\
\hline Vela X-1 \cite{96} & 1.77 $\pm$ 0.08 & 9.56 $\pm$ 0.08 &
0.181282 & 60.7098 & 334.891.
\\
\hline 4U 1608-52 \cite{97} & 1.74 $\pm$ 0.01 & 9.3 $\pm$ 0.10 &
0.232224 & 75.141 & 323.571.
\\
\hline PSR J1903+327 \cite{98} & 1.667 $\pm$ 0.021 & 9.48 $\pm$ 0.03
& 0.222418 & 77.1192 & 346.73.
\\
\hline PSR J1614-2230 \cite{99} & 1.97 $\pm$ 0.04 & 9.69 $\pm$ 0.2 &
0.100997 & 31.6458 & 313.334.
\\
\hline
\end{tabular}
\end{center}
\end{table}
\begin{figure}
\epsfig{file=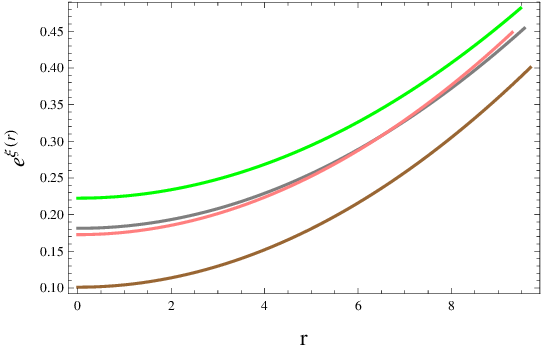,width=.5\linewidth}
\epsfig{file=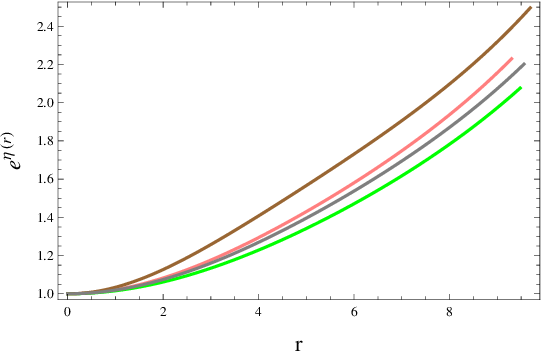,width=.5\linewidth}\caption{Behavior of metric
coefficients for different compact stars.}
\end{figure}

\section{Viable Features of Compact Stars}

Understanding the behavior and properties of CSOs is crucial to
identify their viable features. We analyze the physical parameters
of CSOs and their characteristics through graphs in this section.
The matter density at the center of the CSOs should be maximum and
decrease towards the boundary. The gradient of matter variables
should be zero at the core and exhibit negative trend as it
approaches the surface. Energy constraints should be positive,
preventing exotic or hypothetical forms of matter.

\subsection{Behavior of Matter Contents}

\begin{figure}
\epsfig{file=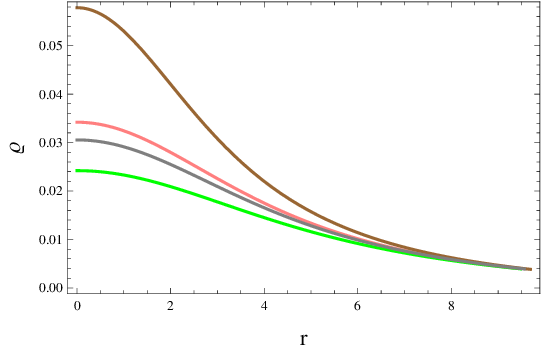,width=.5\linewidth}
\epsfig{file=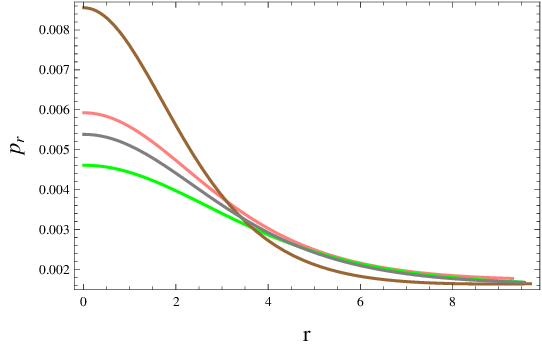,width=.5\linewidth}\center
\epsfig{file=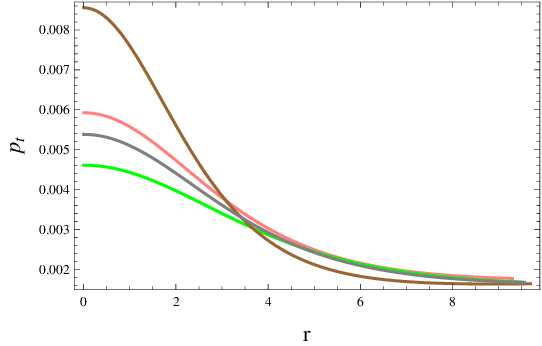,width=.5\linewidth}\caption{Evolution of fluid
parameters.}
\end{figure}

The study of matter contents is significant to comprehend the inner
characteristics of CSOs. These material factors are expected to
reach their peak at a core because of their high density, countering
gravity and preserving the CSOs from collapse. Figures \textbf{2}
and \textbf{3} show the graphical representation of matter contents
and their rates of change for each CSOs. The graphs indicate that
the matter contents reach their highest values at a core before
decreasing, which emphasizes the dense nature of the considered
CSOs. Furthermore, radial pressure in the proposed CSOs exhibits a
consistent decreasing behavior as radial coordinate increases unless
it diminishes at the boundary. Figure \textbf{3} indicates the
highly compact configurations of CSOs in this theory as the rate of
change of matter contents is zero at the core and becomes negative.

Anisotropy $(\Delta=p_t-p_r)$ describes the variation of pressure
across different directions in a system. Anisotropy holds
significance in analyzing the structural configuration of fluid and
their influence on pressure alignments. When anisotropy is positive,
pressure pushes outward, whereas negative anisotropy leads to inward
pressure.The behavior of anisotropy can be observed from Figure
\textbf{2}. Both radial as well as tangential pressures are positive
and $p_t>p_r$, indicating the presence of repulsive force require to
cosmic geometries.
\begin{figure}
\epsfig{file=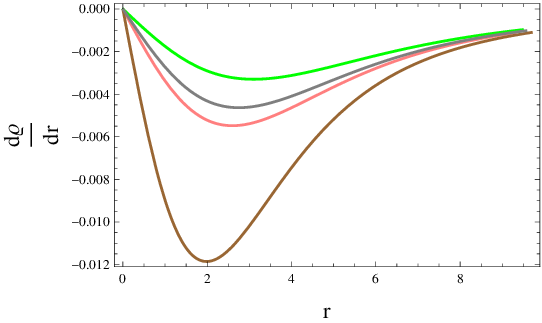,width=.5\linewidth}
\epsfig{file=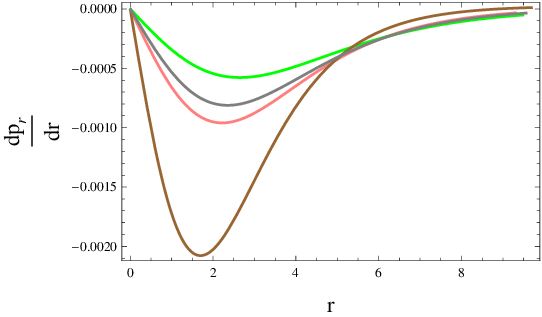,width=.5\linewidth}\center
\epsfig{file=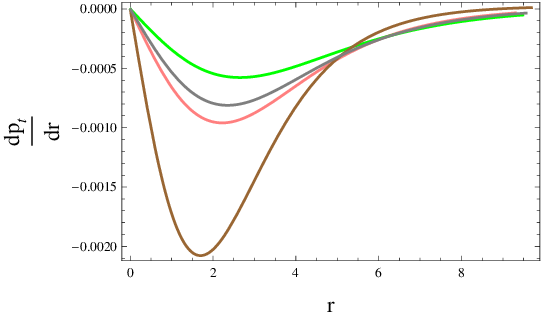,width=.5\linewidth}\caption{Evolution of
gradient of fluid parameters.}
\end{figure}

\subsection{Evolution of Energy Bounds}

Energy conditions are inequalities imposing limits on the EMT.
Different types of energy conditions exist, including dominant
energy condition $(0\leq \varrho\pm p_{r},~ 0\leq \varrho\pm
p_{t})$, null energy condition $(0\leq p_{r}+\varrho, ~0\leq
p_{t}+\varrho)$, weak energy condition $(0\leq p_{r}+\varrho,~ 0\leq
p_{t}+\varrho, ~0\leq \varrho)$ and strong energy condition $(0\leq
p_{r}+\varrho,~ 0\leq p_{t}+\varrho, ~ 0\leq p_{r}+2p_{t}+\varrho)$
which impose different restrictions on the parameters related to
matter. These conditions are significant to determine viable CSOs in
space. All these criteria need to be satisfied for a cosmic
formation to be viable. Figure \textbf{4} demonstrates that with the
inclusion of modified terms, all energy constraints are being
fulfilled which indicate that the matter in the CSOs is normal.
\begin{figure}
\epsfig{file=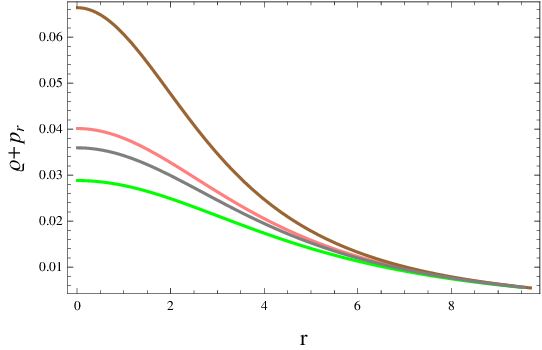,width=.5\linewidth}
\epsfig{file=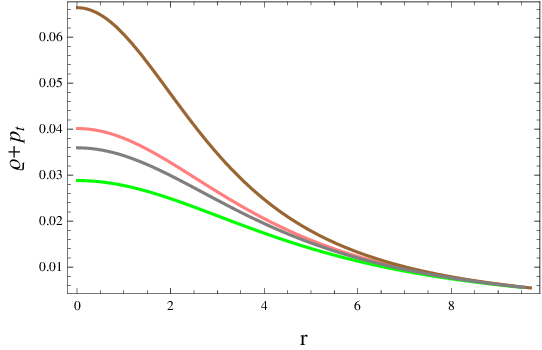,width=.5\linewidth}
\epsfig{file=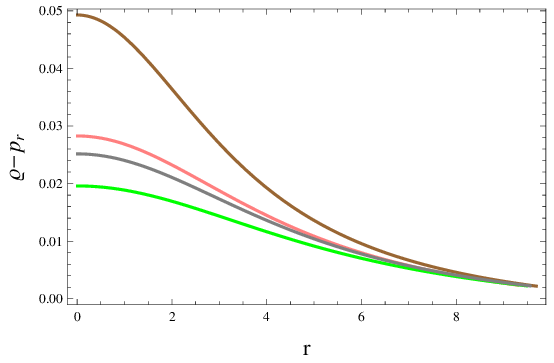,width=.5\linewidth}
\epsfig{file=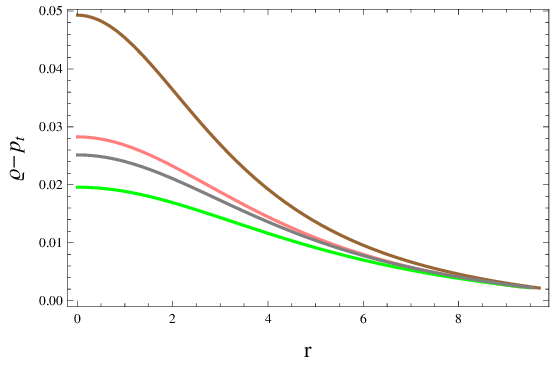,width=.5\linewidth}\caption{Graphs of energy
bounds for different compact stars.}
\end{figure}

\subsection{Graphical Analysis of State Parameters}

In this analysis, we examine the radial
$(\omega_{r}=\frac{p_{r}}{\varrho})$ and tangential
$(\omega_{t}=\frac{p_{t}}{\varrho})$ EoS parameters to establish a
connection among pressure and density in physical contexts. To
ensure that our model is physically feasible, these parameters
should be in the range of [0,1]. Using Eqs.(\ref{13a})-(\ref{15a}),
we have
\begin{eqnarray}\nonumber
\omega_{r}&=&\frac{(b+2r^{2})(b-c+3r^{2})+(b(2b+c)+6b
r^{2}+6r^{4})k}{b^{2}(2k-3)-2r^{2}(c-4c
k+3r^{2}(1+k))+b(c(7k-3)-r^{2}(2k+7))},
\\\nonumber
\omega_{t}&=&\frac{(b+2r^{2})(b-c+3r^{2})+(b(2b+c)+6b
r^{2}+6r^{4})k}{b^{2}(2k-3)-2r^{2}(c-4c
k+3r^{2}(1+k))+b(c(7k-3)-r^{2}(2k+7))}.
\end{eqnarray}
Figure \textbf{5} determine viable CSOs as the EoS parameters
satisfy the limits.
\begin{figure}
\epsfig{file=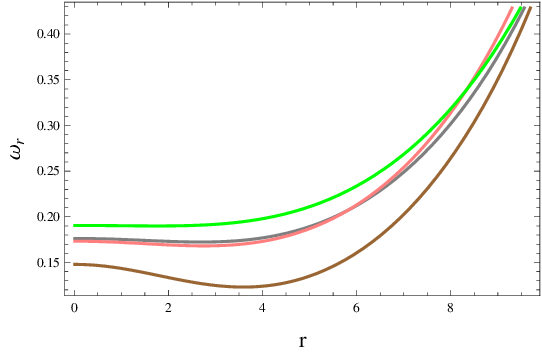,width=.5\linewidth}
\epsfig{file=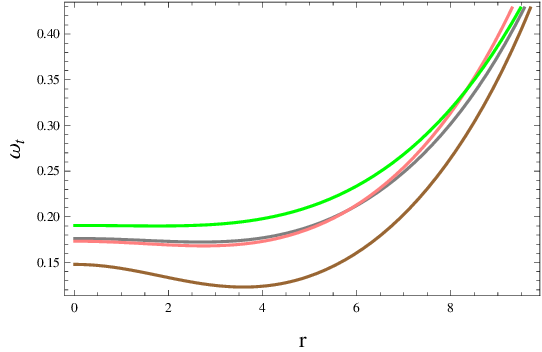,width=.5\linewidth}\caption{Graphs of EoS
parameter for different compact stars.}
\end{figure}

\subsection{Study of Various Physical Aspects}

The mass of CSO is expressed as
\begin{equation}\label{19}
M=4\pi\int^{{R}}_{0} r^{2}\varrho dr.
\end{equation}
Figure \textbf{6} illustrate that the mass function increases
monotonically and become zero at center, indicating no singularities
or irregularities in the mass distribution. Compactness
$(u=\frac{M}{r})$ a key element in determining the feasibility of
CSOs, provides information on how mass is distributed in relation to
the radius of a CSO and how much mass is concentrated. To ensure a
viable distribution of mass, Buchdahl's condition dictates that for
a viable CSO as $u<4/9$ \cite{100}. Surface redshift is a light
emitted from star's surface caused by the gravitational redshift.
This parameter is crucial for studying CSOs, defined as
\begin{equation}\label{19a}
Z_s =- 1+\frac{1}{\sqrt{1-2u}}.
\end{equation}
In the case where there is anisotropic configuration, surface
redshift needs to meet a particular requirement of $Z_{s}<5.211$ for
feasible CSOs \cite{101}. Figure \textbf{7} demonstrates that graph
of both compactness and redshift functions continuously increase and
fulfill the necessary feasibility criteria.
\begin{figure}\center
\epsfig{file=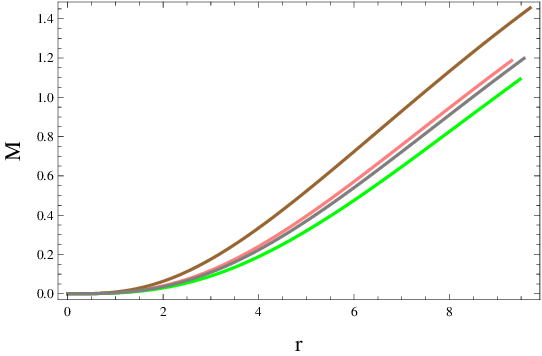,width=.5\linewidth}\caption{Plot of mass
function.}
\end{figure}
\begin{figure}
\epsfig{file=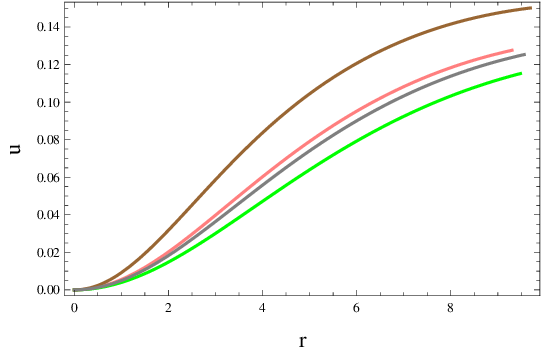,width=.5\linewidth}
\epsfig{file=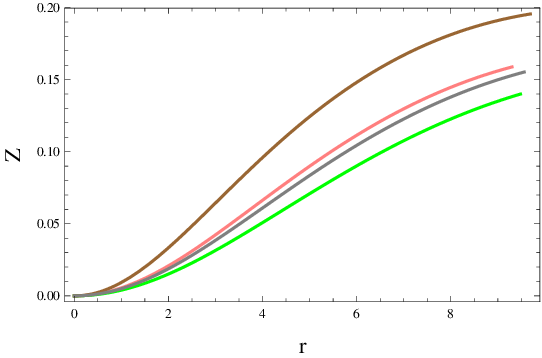,width=.5\linewidth}\caption{Graph of
compactness and redshift functions.}
\end{figure}

\subsection{Analysis of Different Forces}

The TOV equation refers an equilibrium state of static spherical
symmetric objects \cite{102}, defined as
\begin{equation}\label{20}
\frac{M_{G}(r)}{r^{2}}(\varrho+p_{r})e^\frac{\xi-\eta}{2}+
p_{r}'-\frac{2\Delta}{r},
\end{equation}
where
\begin{equation}\nonumber
M_{G}(r)=4\pi \int
(\mathbb{T}^{0}_{0}-\mathbb{T}^{1}_{1}-\mathbb{T}^{2}_{2}-\mathbb{T}^{3}_{3})r^{2}
e^{\frac{\xi+\eta}{2}}dr.
\end{equation}
Its solution yields
\begin{equation}\nonumber
M_{G}(r)=\frac{1}{2}r^{2} e^{\frac{\eta-\xi}{2}}\xi'.
\end{equation}
Equation (\ref{20a}) turns out to be
\begin{equation}\label{21}
\frac{1}{2}\xi'(\varrho+p_{r})+ p_{r}'-\frac{2\Delta}{r}=0.
\end{equation}
This equation demonstrates the influence of gravitational force
($F_{g}=\frac{\xi'(\varrho+p_{r})}{2}$), hydrostatic force
($F_{h}=p_{r}'$) and anisotropic force ($F_{a}=\frac{2\Delta}{r}$).
The graphical representation of TOV equation is given in Figure
\textbf{8} which demonstrates that the stellar objects under
consideration are in the state of equilibrium.
\begin{figure}\center
\epsfig{file=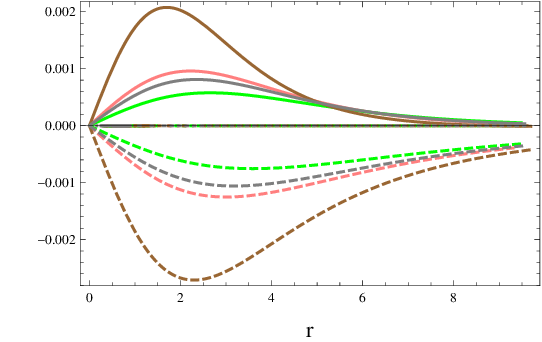,width=.5\linewidth}\caption{Plot of TOV
equation.}
\end{figure}

\section{Stability Analysis}

Scientists examined the constraints that demonstrate the stability
against different types of oscillations. To explore the stability of
CSOs, researchers employ the sound speed method and adiabatic index,
offering valuable insights into the structural soundness and
resilience of celestial objects.

\subsection{Sound Speed}

To analyze the stability of CSOs, the causality principle can be
considered which states that no information can exceed the speed of
light. It is necessary for the radial ($u_r =
\frac{dp_r}{d\varrho}$) and tangential $(u_t =
\frac{dp_t}{d\varrho}$) velocities of sound to be in the range of 0
and 1 \cite{103} for stable configuration. Figure \textbf{9} shows
that the sound speed components lie in the required limit. The
Herrera cracking states that CSOs are stable when the difference
between $u_{st}$ and $u_{st}$ falls in the range of 0 to 1.
According to Figure \textbf{9} the difference of radial and
tangential velocities of sound lie between 0 and 1. Thus, the
$f(\mathbb{Q},\mathbb{T})$ gravity provides evidence for the
presence of physically stable CSOs.
\begin{figure}
\epsfig{file=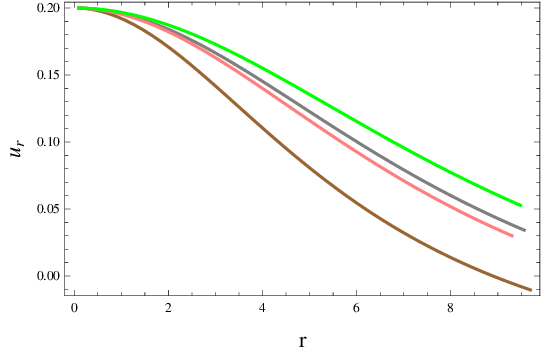,width=.5\linewidth}
\epsfig{file=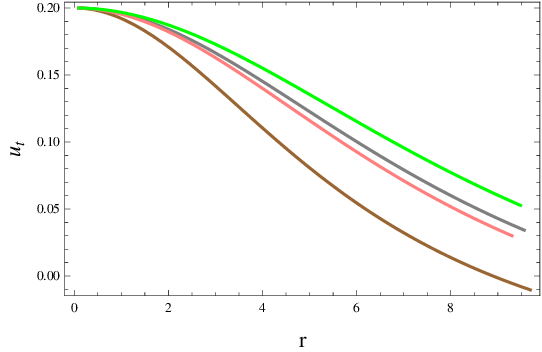,width=.5\linewidth}\caption{Plots of causality
condition.}
\end{figure}

\subsection{Adiabatic Index}
\begin{figure}
\epsfig{file=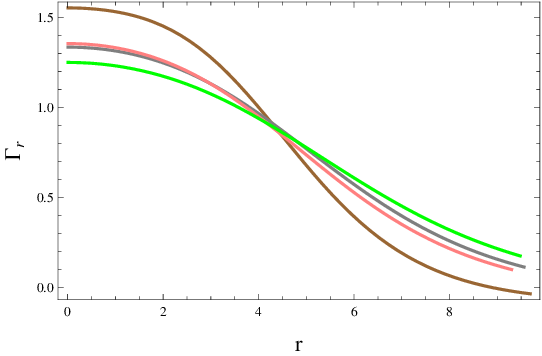,width=.5\linewidth}
\epsfig{file=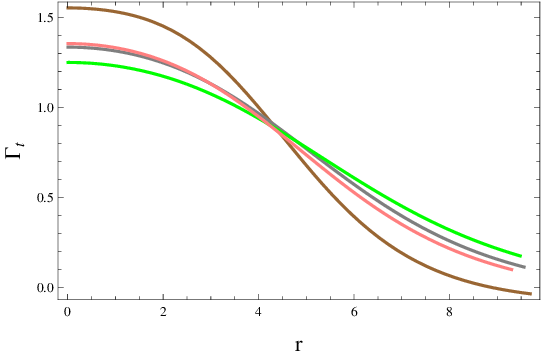,width=.5\linewidth}\caption{Stability analysis
by adiabatic index.}
\end{figure}

An adiabatic index determines the change in pressure corresponding
to density, defined as
\begin{eqnarray}\label{22}
\Gamma_{r}=\frac{\varrho+p_{r}}{p_{r}}u_{r},\quad
\Gamma_{t}=\frac{\varrho+p_{t}}{p_{t}}u_{t}.
\end{eqnarray}
In context of CSOs, adiabatic index provides information about how
pressure and density variations are linked in the star. For
stability by adiabatic process, the CSOs should satisfy the
particular constraint as $\Gamma>4/3$ \cite{104}. Using
Eqs.(\ref{13a})-(\ref{15a}), the above equation becomes
\begin{eqnarray}\nonumber
\Gamma_{r}&=&-\bigg[2(b+2c)(b+r^{2})(2k-1)(b+2r^{2}-2b
k)\bigg]\bigg[((b+2r^{2})(b-c+3r^{2})
\\\label{23}
&+&(b(2b+c)+6b r^{2}+6r^{4})k)(5b(2k-1)+2r^{2}(4k-1))\bigg]^{-1},
\\\nonumber
\Gamma_{t}&=&-\bigg[2(b+2c)(b+r^{2})(2k-1)(b+2r^{2}-2b
k)\bigg]\bigg[((b+2r^{2})(b-c+3r^{2})
\\\label{24}
&+&(b(2b+c)+6b r^{2}+6r^{4})k)(5b(2k-1)+2r^{2}(4k-1))\bigg]^{-1}.
\end{eqnarray}
Figure \textbf{10} shows that the CSOs under consideration are
stable as they fulfill necessary condition. Thus, physically stable
CSOs exist in modified $f(\mathbb{Q},\mathbb{T})$ gravity.

\section{Conclusions}

Modified gravity theories such as $f(\mathbb{Q},\mathbb{T})$ gravity
have received considerable interest in recent years as alternate
explanations to GR. This theory suggests alterations to the
fundamental equations of gravity with the aim of addressing numerous
unresolved inquiries in the fields of cosmology and astrophysics. A
crucial method for assessing and limiting the validity of this
modified gravity theory is by conducting astrophysical observations.
This modified theory introduces extra independent variables that
describe deviations from GR. This modified proposal aims to enhance
our understanding of the principles that regulate the actions of
matter and energy in the cosmos, providing opportunities to
investigate novel phenomena that may not be accounted for in GR. The
inclusion of correction terms emphasize the significant results,
imposing a large impact on the geometric feasibility. The
$f(\mathbb{Q},\mathbb{T})$ theory is used as a mathematical tool to
examine complex elements of gravitational dynamics on a large scale.
The motivation behind the investigation of this theory involves
analyzing its theoretical implications, its consistency with
empirical facts and its importance in cosmological contexts.

The CSOs characterized by anisotropic matter exhibiting a stable and
theoretically viable configuration is the subject of the
$f(\mathbb{Q},\mathbb{T})$ gravity. Using non-metricity and the
matter source, our research aims to determine a theoretically viable
gravitational field equations. The hypothesis is based on the
expectation that the introduction of additional terms provide novel
insights into the behavior of matter and geometry in extreme
gravitational environments. The proposed CSOs were confirmed to be
stable under certain conditions. The main findings are summarized
below.
\begin{itemize}
\item
Metric coefficients ensure a smooth singularity free spacetime
(Figure \textbf{1}).
\item
The regular and maximal matter contents at origin of CSOs shows a
stable core that reduces towards the boundary, satisfying viability
of the CSOs (Figure \textbf{2}).
\item
A dense profile of the CSOs satisfy their viability as they exhibit
negative gradient of matter contents (Figure \textbf{3}).
\item
The energy constraints ensure the viability of CSOs (Figure
\textbf{4}).
\item
The EoS parameters ensure the viability of CSOs (Figure \textbf{5}).
\item
The mass function increases with radial distance and vanishes at the
core limit to ensure the viability of CSOs (Figure \textbf{6}).
\item
The compactness factor and surface redshift ensure the viability of
CSOs (Figure \textbf{7}).
\item
The TOV equation ensure the equilibrium state of CSOs (Figure
\textbf{8}).
\item
All CSOs are stable in the $f(\mathbb{Q},\mathbb{T})$ theory as a
stability limits are satisfied (Figures \textbf{9}-\textbf{10}).
\end{itemize}

In this research, we have examined that CSOs characterized by
anisotropic matter in $f(\mathbb{Q},\mathbb{T})$ theory exhibit
theoretical viable and stable configurations. Specifically, we aim
to explore whether the inclusion of non-metricity and the matter
source in the gravitational field equations leads to theoretical
viable solutions for anisotropic stellar objects. Our hypothesis is
based on the expectation that incorporating these additional terms
in the theory provide novel insights into the behavior of matter and
geometry in extreme gravitational environments. Furthermore, we have
predicted that graphical analysis of various physical quantities
ensure the viability of the CSOs in the proposed theoretical
framework. Additionally, we have confirmed the stability of the
proposed CSOs under certain conditions. Overall, we anticipate that
our research conclude that the CSOs analyzed in the
$f(\mathbb{Q},\mathbb{T})$ theory framework are both theoretical
viable and stable, thereby contributing to our understanding of
stellar structures and alternative theories of gravity.

Notably, we observed that all parameters reach their maximum values
when compared to GR \cite{105}-\cite{107} and other modified
gravitational theories \cite{108}. In the realm of $f(\mathbb{R})$
theory, the results indicate the instability of the Her X-1 CSO
associated with the second gravity model due to the limited range
satisfied by the physical quantities \cite{109}. In
$f(\mathbb{G},\mathbb{T})$ theory, it is analyzed that the evolution
of CSOs (SAXJ1808.4-3658 and 4U 1820-30) is supported by all three
gravity models while for Her X-1, there are some restrictions
asserted by the second model to be completely physically viable
\cite{110}. Furthermore, in the framework of
$f(\mathbb{R},\mathbb{T}^{2})$ theory, it is found that CSOs are
neither physically viable nor stable at the center \cite{111}. In
light of these findings, it can be concluded that all considered
CSOs exhibit both physical viability and stability at the center in
this modified theory. Consequently, our results suggest that viable
and stable CSOs can exist in this modified theory. Therefore, we
conclude that the solutions we have obtained are physically valid,
providing stable and viable structures for anisotropic CSOs.

\vspace{0.25cm}

\section*{Appendix A: Calculation of $\mathbb{Q}$}
\renewcommand{\theequation}{A\arabic{equation}}
\setcounter{equation}{0}

Using Eqs.(\ref{18}) and (\ref{19}), we have
\begin{eqnarray}\label{A1}
\mathbb{Q} &\equiv& -\mathrm{g}^{\i \nu} (\mathbb{L}^{\chi}_{~\tau
\i}\mathbb{L}^{\tau}_{~\nu\chi} - \mathbb{L}^{\chi}_{~\tau
\chi}\mathbb{L}^{\tau}_{~\i\nu}),
\\\label{A2}
\mathbb{L}^{\chi}_{~\tau \i}&=&-\frac{1}{2} \mathrm{g}^{\chi
\eta}(\mathbb{Q}_{\i \tau \eta} +\mathbb{Q}_{\tau \eta
\i}-\mathbb{Q}_{\eta\i\tau}),
\\\label{A3}
\mathbb{L}^{\tau}_{~\nu \chi}&=&-\frac{1}{2} \mathrm{g}^{\tau
\eta}(\mathbb{Q}_{\chi \nu \eta} +\mathbb{Q}_{\nu \eta
\chi}-\mathbb{Q}_{\eta \chi \nu}),
\\\label{A4}
\mathbb{L}^{\chi}_{~\tau \i} &=& -\frac{1}{2} \mathrm{g}^{\chi
\eta}(\mathbb{Q}_{\chi \tau \eta} +\mathbb{Q}_{\tau \eta
\chi}-\mathbb{Q}_{\eta \chi \tau}),
\\\label{A5}
&=&-\frac{1}{2}(\bar{\mathbb{Q}}_{\tau}
+\mathbb{Q}_{\tau}-\bar{\mathbb{Q}}_{\tau})=-\frac{1}{2}
\mathbb{Q}_{\tau},
\\\label{A6}
\mathbb{L}^{\tau}_{~\i \nu}&=& -\frac{1}{2}\mathrm{g}^{\tau
\eta}(\mathbb{Q}_{\nu \i \eta} +\mathbb{Q}_{\i \eta
\nu}-\mathbb{Q}_{\eta \nu \i}).
\end{eqnarray}
Thus, we have
\begin{eqnarray}\nonumber
-\mathrm{g}^{\i \nu}\mathbb{L}^{\chi}_{~\tau \i}
\mathbb{L}^{\tau}_{~\nu \chi}&=& -\frac{1}{4}\mathrm{g}^{\i
\nu}\mathrm{g}^{\chi\eta} \mathrm{g}^{\tau \eta} (\mathbb{Q}_{\i
\tau \eta}+\mathbb{Q}_{\tau \eta \i} -\mathbb{Q}_{\eta \i \tau})
\\\label{A7}
&\times&(\mathbb{Q}_{\chi \nu \eta}+\mathbb{Q}_{\nu \eta \chi}
-\mathbb{Q}_{\eta \chi \nu}),
\\\nonumber
&=&-\frac{1}{4}(\mathbb{Q}^{\nu \eta \chi}+\mathbb{Q}^{\eta \chi
\nu} -\mathbb{Q}^{\chi \nu \eta})
\\\label{A8}
&\times&(\mathbb{Q}_{\chi \nu \eta}+\mathbb{Q}_{\nu \eta \chi}
-\mathbb{Q}_{\eta \chi \nu}),
\\\label{A9}
&=&-\frac{1}{4}(2\mathbb{Q}^{\nu \eta \chi}\mathbb{Q}_{\eta \chi
\nu} - \mathbb{Q}^{\nu \eta \chi}\mathbb{Q}_{\nu \eta \chi}),
\\\label{A10}
\mathrm{g}^{\i \nu}\mathbb{L}^{\chi}_{~\tau
\chi}\mathbb{L}^{\tau}_{~\i \nu}&=& \frac{1}{4}\mathrm{g}^{\i
\nu}\mathrm{g}^{\tau \eta}\mathbb{Q}_{\eta} (\mathbb{Q}_{\nu \i
\eta}+\mathbb{Q}_{\i \eta \nu} -\mathbb{Q}_{\eta \nu \i}),
\\\label{A11}
&=&\frac{1}{4}\mathbb{Q}^{\eta}(2\bar{\mathbb{Q}_{\eta}}-
\mathbb{Q}_{\eta}),
\\\nonumber
\mathbb{Q}&=& -\frac{1}{4}(\mathbb{Q}^{\chi \nu
\eta}\mathbb{Q}_{\chi \nu \eta} +2\mathbb{Q}^{\chi \nu \eta
\chi}\mathbb{Q}_{\eta \chi \nu}
\\\label{A12}
&-&2\mathbb{Q}^{\eta}\bar{\mathbb{Q}_{\eta}}+\mathbb{Q}^{\eta}\mathbb{Q}_{\eta}).
\end{eqnarray}
Thus, Eq.(\ref{25}) yields
\begin{eqnarray}\nonumber
\mathbb{P}^{\chi \i \nu}&=&\frac{1}{4}[-\mathbb{Q}^{\chi \i \nu}
+\mathbb{Q}^{\i \chi \nu}+\mathbb{Q}^{\nu \chi \i}
+\mathbb{Q}^{\chi}\mathrm{g}^{\i \nu}
\\\label{A13}
&-&\bar{\mathbb{Q}^{\chi}}\mathrm{g}^{\i \nu}-\frac{1}{2}
(\mathrm{g}^{\chi \i} \mathbb{Q}^{\nu}+\mathrm{g}^{\chi
\nu}\mathbb{Q}^{\i})],
\\\nonumber
-\mathbb{Q}_{\chi \i \nu}\mathbb{P}^{\chi \i \nu} &=&
-\frac{1}{4}[-\mathbb{Q}_{\chi \i \nu}\mathbb{Q}^{\chi \i \nu}
\\\nonumber
&+&\mathbb{Q}_{\chi \i \nu}\mathbb{Q}^{\i \chi \nu} +\mathbb{Q}^{\nu
\chi \i} \mathbb{Q}_{\chi \i \nu}+\mathbb{Q}_{\chi \i \nu}
\mathbb{Q}^{\chi}\mathrm{g}^{\i \nu}
\\\label{A14}
&-&\mathbb{Q}_{\chi \i \nu}\bar{\mathbb{Q}^{\chi}} \mathrm{g}^{\i
\nu} -\frac{1}{2}\mathbb{Q}_{\chi \i \nu}(\mathrm{g}^{\chi \i}
\mathbb{Q}^{\nu} +\mathrm{g}^{\chi \nu}\mathbb{Q}^{\i})],
\\\label{A15}
&=& -\frac{1}{4}(-\mathbb{Q}_{\chi \i \nu}\mathbb{Q}^{\chi \i \nu}
+2\mathbb{Q}_{\chi \i \nu}\mathbb{Q}^{\i \chi \nu}+\mathbb{Q}^{\chi}
\mathbb{Q}_{\chi}-2\tilde{\mathbb{Q}^{\chi}}\mathbb{Q}_{\chi}),
\\\label{A16}
&=& \mathbb{Q}.
\end{eqnarray}

\section*{Appendix B: Variation of $\mathbb{Q}$}
\renewcommand{\theequation}{B\arabic{equation}}
\setcounter{equation}{0}

The non-metricity tensors are defined as
\begin{eqnarray}\label{B1}
\mathbb{Q}_{\chi \i \nu}&=&\nabla_{\chi}\mathrm{g}_{\i \nu},
\\\label{B2}
\mathbb{Q}^{\chi}~_{\i \nu}&=&\mathrm{g}^{\chi \tau}
\mathbb{Q}_{\tau \i \nu} =\mathrm{g}^{\chi
\tau}\nabla_{\tau}\mathrm{g}_{\i \nu} =\nabla^{\chi}\mathrm{g}_{\i
\nu},
\\\label{B3}
\mathbb{Q}_{\chi~~\nu}^{~~\i}&=&\mathrm{g}^{\i\eta}\mathbb{Q}_{\chi
\eta \nu} =\mathrm{g}^{\i\eta}\nabla_{\chi}\mathrm{g}_{\eta \nu}
=-\mathrm{g}_{\i\eta}\nabla_{\chi}\mathrm{g}^{\i\eta},
\\\label{B4}
\mathbb{Q}_{\chi\eta}^{~~\nu} &=& \mathrm{g}^{\nu
\eta}\mathbb{Q}_{\chi \i \eta} =\mathrm{g}^{\nu
\eta}\nabla_{\chi}\mathrm{g}_{\i\eta}
=-\mathrm{g}_{\i\eta}\nabla_{\chi}\mathrm{g}^{\nu \eta},
\\\label{B5}
\mathbb{Q}^{\chi \i}_{~~\nu}&=& \mathrm{g}^{\i\eta}\mathrm{g}^{\chi
\tau}\nabla_{\tau}\mathrm{g} _{\eta \nu}
=\mathrm{g}^{\i\eta}\nabla^{\chi}\mathrm{g}_{\nu \eta}
=-\mathrm{g}_{\eta \nu}\nabla^{\chi}\mathrm{g}^{\i\eta},
\\\label{B6}
\mathbb{Q}^ {\chi~~\nu} _{~\i} &=& \mathrm{g}^{\nu
\eta}\mathrm{g}^{\chi \tau}\nabla_{\tau}\mathrm{g} _{\i\eta}
=\mathrm{g}^{\nu \eta}\nabla^{\chi}\mathrm{g}_{\i\eta}
=-\mathrm{g}_{\i\eta}\nabla^{\chi}\mathrm{g}^{\nu \eta},
\\\label{B7}
\mathbb{Q}_{\chi}^{~~\i \nu}&=& \mathrm{g}^{\i\eta}\mathrm{g}^{\nu
\tau}\nabla_{\chi}\mathrm{g} _{\eta \tau}
=-\mathrm{g}^{\i\eta}\mathrm{g}_{\eta \tau}\nabla_{\chi}
\mathrm{g}^{\nu \eta} =-\nabla_{\chi}\mathrm{g}^{\i \nu}.
\end{eqnarray}
By using Eq.(\ref{A12}), we have
\begin{eqnarray}\label{B8}
\delta \mathbb{Q} &=&-\frac{1}{4} \delta(-\mathbb{Q}^{\chi \nu \eta}
\mathbb{Q}_{\chi \nu \eta}+2\mathbb{Q}^{\chi \nu \eta}
\mathbb{Q}_{\eta \chi \nu}-2\mathbb{Q}^{\eta}
\bar{\mathbb{Q}_{\eta}}+\mathbb{Q}^{\eta}\mathbb{Q}_{\eta}),
\\\nonumber
&=&-\frac{1}{4}(-\delta \mathbb{Q}^{\chi \nu \eta} \mathbb{Q}_{\chi
\nu \eta} - \mathbb{Q}^{\chi \nu \eta}\delta \mathbb{Q}_{\chi \nu
\eta} + 2\delta \mathbb{Q}_{\chi \nu \eta}\mathbb{Q}^{\eta \chi \nu}
\\\label{B9}
&+& 2 \mathbb{Q}^{\chi \nu \eta}\delta \mathbb{Q}_{\eta \chi
\nu}-2\delta \mathbb{Q}^{\eta}\bar{\mathbb{Q}_{\eta}}+\delta
\mathbb{Q}^{\eta}\mathbb{Q}_{\eta}-2 \mathbb{Q}^{\eta}\delta \bar
{\mathbb{Q}_{\eta}} + \mathbb{Q}^{\eta}\delta \mathbb{Q}_{\eta}),
\\\nonumber
&=&-\frac{1}{4}[\mathbb{Q}_{\chi \nu \eta}\nabla ^{\chi}\delta
\mathrm{g}^{\nu \eta}-\mathbb{Q}^{\chi\nu\varsigma}
\nabla_{\chi}\delta \mathrm{g}_{\nu \eta}-2\mathbb{Q}_{\eta \chi
\nu} \nabla^{\chi}\delta \mathrm{g}^{\nu \eta}
\\\nonumber
&+&2\mathbb{Q}^{\chi \nu \eta}\nabla_{\eta}\delta \mathrm{g}_{\chi
\nu}+ 2\bar{\mathbb{Q}_{\eta}}\mathrm{g}^{\i \nu}\nabla
^{\eta}\delta \mathrm{g}_{\i
\nu}+2\mathbb{Q}^{\eta}\nabla^{\tau}\delta \mathrm{g}_{\eta \tau}
\\\nonumber
&+&2\bar{\mathbb{Q}_{\eta}} \mathrm{g}_{\i \nu}\nabla^{\eta}\delta
\mathrm{g}^{\i \nu}-\mathbb{Q}_{\eta}\nabla^{\tau}\mathrm{g} ^{\i
\nu}\delta \mathrm{g}_{\i \nu}-\mathbb{Q}_{\eta}\mathrm{g}_{\i
\nu}\nabla^{\eta}\delta \mathrm{g}^{\i \nu}
\\\label{B10}
&-&\mathbb{Q}_{\eta} \mathrm{g}^{\i \nu} \nabla_{\eta} \delta
\mathrm{g}_{\i \nu} -\mathbb{Q}^{\eta}\mathrm{g}_{\i
\nu}\nabla_{\eta}\delta \mathrm{g}_{\i \nu}].
\end{eqnarray}
The above equation is simplified by using the following relations.
\begin{eqnarray}\label{B11}
\delta \mathrm{g}_{\i \nu}&=&-\mathrm{g}_{\i \chi} \delta
\mathrm{g}^{\chi \tau}\mathrm{g}_{\tau \nu}-\mathbb{Q}^ {\chi \nu
\eta} \nabla_{\chi}\delta \mathrm{g}_{\nu \eta},
\\\label{B12}
&=&-\mathbb{Q}^{\chi \nu \eta} \nabla_{\chi}(-\mathrm{g}_{\nu
\i}\delta \mathrm{g}^{\i \tau}\mathrm{g}_{\tau \eta}),
\\\label{B12a}
&=&2\mathbb{Q}_{~~\eta}^{\chi \nu}\mathbb{Q}_{\chi \nu \i}\delta
\mathrm{g}^{\i\theta} + \mathbb{Q}_{\chi \tau
\eta}\nabla^{\chi}\mathrm{g}^{\i\eta},
\\\label{B13}
&=&2\mathbb{Q}_{~~\nu}^{\chi \tau}\mathbb{Q}_{\chi \tau \nu}\delta
\mathrm{g}^{\i \nu}+\mathbb{Q}_{\chi \nu \eta}\nabla^{\chi}
\mathrm{g}^{\nu \eta},
\\\label{B14}
2\mathbb{Q}^{\chi \nu \eta}\nabla_{\eta}\delta \mathrm{g}_{\chi
\nu}&=& -4\mathbb{Q}_{\i}^{~\tau \eta}\mathbb{Q}_{\eta \tau \nu}
\delta \mathrm{g}^{\i \nu}-2 \mathbb{Q}_{\nu \eta
\chi}\nabla^{\chi}\delta \mathrm{g}^{\nu \eta},
\\\nonumber
-2\mathbb{Q}^{\eta} \nabla^{\tau} \delta \mathrm{g}_{\eta
\tau}&=&2\mathbb{Q}^{\chi} \mathbb{Q}_{\nu \chi \i} \delta
\mathrm{g}^{\i \nu}+ 2\mathbb{Q}_{\i}\bar{\mathbb{Q}_{\nu}} \delta
\mathrm{g}^{\i \nu}
\\\label{B15}
&+&2\mathbb{Q}_{\nu}\mathrm{g}_{\chi \eta}\nabla^{\chi}
\mathrm{g}^{\nu \eta}.
\end{eqnarray}
Thus, we have
\begin{equation}\label{B16}
\delta\mathbb{Q}=2\mathbb{P}_{\chi \nu \eta}\nabla^{\chi}\delta
\mathrm{g}^{\nu \eta}-(~\mathbb{P}_{\i \chi \tau}\mathbb{Q}_{\nu}
^{~\chi \tau}-2 \mathbb{P}_{\chi \tau \nu}\mathbb{Q}^{\chi \tau}
_{~\nu})\delta \mathrm{g}^{\i \nu}.
\end{equation}

\textbf{Data Availability Statement:} No new data were generated or
analyzed in support of this research.

\end{document}